\documentclass[12pt,preprint]{aastex}
\def\asca{{\it ASCA}}
\def\lum{{ergs s$^{-1}$}}
\def\etal{{\it et al.}}
\def\apj{{Astrophys. J.}}
\def\apjs{{Astorophys. J. Suppl.}}

\def\aanda{{Astron. \& Astrophys.}}
\def\aandas{{Astron. \& Astrophys. Suppl.}}
\def\mnras{{Mon. Not. R. Astron. Soc.}}
\def\araa{{Annu. Rev. Astron. Astrophys.}}

\def\pasj{{Publ. Astron. Soc. Japan}}

\lefthead{Akimoto et al.}
\righthead{Temperature and Abundance Map of A3558}

\begin{document}

\title{
Spatial Distributions of A3558 in the Core Region of the Shapley Supercluster
}
\author{F. Akimoto, K. Kondou, A. Furuzawa, Y. Tawara and K. Yamashita}
\affil{Department of Physics, Nagoya University}
\authoraddr{Furo-cho, Chikusa-ku, Nagoya 464-8602, Japan}

\begin{abstract}

The core region of the Shapley Supercluster is dominated by three rich
Abell clusters and two poor clusters. Since these member clusters are
expected to be evolving rapidly in comparison to nonmember clusters
because of the high merging rate, it is important to study the member
clusters for understanding of the cluster evolution. Since the spatial
distributions of gas temperature and metal abundance in each member
cluster provide us with information on the interactions and motions of
member clusters, they are useful for understanding their dynamics.
From the results of eight $ASCA$ pointing observations (total $\sim$
300 ksec) of the core region, we obtained parameters of gas
temperature, metal abundance, and X-ray luminosity for five member
clusters and found that they are similar to the other field clusters
not belonging to superclusters observed with $ASCA$. This result and
the mean gravitational mass density of the core region indicate that
the members are growing in the same way as the non-member clusters,
and the core of the supercluster is just on the way to contraction.
Based on analyses of detailed spatial structures with a 4$^\prime$
$\times$ 4$^\prime$ scale, the two poor clusters show nearly isotropic
temperature distributions, while the three Abell clusters are
asymmetric. A3558 was analyzed with a 2$^\prime$ $\times$ 2$^\prime$
scale, owing to the statistical advantage, and it was revealed that
A3558 has clear asymmetric distributions of gas temperature and X-ray
surface brightness. This is thought to be caused by cluster-cluster
mergings and/or group infallings. A metal-rich region with the size of
$\sim$ 320 $h_{50}^{-1}$ kpc was also found to the southeast, $\sim$
12$^\prime$ away from the cluster center of A3558. It is expected that
either a remnant of a merged core has been left after a major merging
or a group of galaxies has been recently infalling. Thus, the high
dynamical activity of A3558 is proved.

\end{abstract}


\section{Introduction}

Superclusters have the highest number density of galaxy clusters in
the universe. Compared to field galaxy clusters, the member clusters
in a supercluster are thought to merge easily with each other, or with
galaxies and groups of galaxies within the potential well of the
supercluster. Thus, the dynamical state of supercluster members is a
clue to interpreting the evolution of large-scale structures in the
universe.

The Shapley Supercluster, SCL 124 \citep{einasto97}, has 25 Abell
clusters of galaxies within a radius of 50 $h_{50}^{-1}$ Mpc and is
the region with the highest known number density of member clusters. 
In particular, the core region of the Shapley Supercluster is defined
to be a region lying east to west over 15 $h_{50}^{-1}$ Mpc
\citep{bard98b} that contains five clusters of galaxies. 
Easternmost are A3562, SC 1329$-$313, SC 1327$-$312, A3558 is in the
center position, and westernmost is A3556, as shown in Figure 1.

Since the core region was observed in the optical wavelength by 
\citet{shapley30}, many multiwavelength observations have been
done. The radial velocity and distribution of the galaxies has been
investigated with optical observations in detail
\citep{bard98b, bard00}. X-ray observations have also been done 
by various satellites; $GINGA$ \citep{day91}, $Einstein$
\citep{raycha91, breen94}, $ROSAT$ \citep{ettori97, ettori00}, $ASCA$
(\citet{marke97} for A3558, \citet{hana99} for five core members). 
Temperature distributions of A3558 and SC 1329$-$313 were investigated
by \citet{marke97} and \citet{hana99}, respectively, and they reported
asymmetric distributions. The total gravitational mass of the core
region is estimated to be 10$^{15}$ $\sim$ 10$^{16}$ $h_{50}^{-1}$
M$_{\odot}$ from galaxy radial velocity dispersion
\citep{metcalfe94} and X-ray observation \citep{raycha91, ettori97}.

The central galaxy cluster in the core region of the Shapley
Supercluster, A3558 is one of the richest clusters of galaxies. 
Observational results in X-ray band of A3558 estimate the gas
temperature of the overall region to be 3 $\sim$ 4 keV \citep{bard96}
with $ROSAT$, and 5 $\sim$ 6 keV \citep{day91, marke97, hana99} with
$GINGA$ and $ASCA$. There is a difference among the satellites because
of the limitation of the observable energy band. The metal abundance
and gravitational mass of A3558 are estimated to be $\sim$ 0.3 solar
\citep{day91, bard96, hana99} and 3 $\sim$ 6 $\times$ 10$^{14}$
$h_{50}^{-1}$ M$_{\odot}$ \citep{bard96, ettori97}, respectively. The
temperature and abundance of the whole region are typical values for a
galaxy cluster.

A3558 has a galaxy density distribution elongated in the
northwest--southeast direction, and many groups and clusters of
galaxies are aligned along this major axis \citep{bard94}. The X-ray
surface brightness distribution also elongates in the same direction
\citep{bard96}. Therefore, in this core region of the Shapley
Supercluster, the infall of matter along this axis is thought to be
dominant. From these characteristics and the two-dimensional
distributions of gas temperature and metal abundance obtained by
$ASCA$ observations, we discuss the dynamical structure of A3558.

This paper comprises six sections. In section 2, the details of the
$ASCA$ observations and analysis method are described. We present the
results of the $ASCA$ data analysis for A3558 in section 3 and for
other member clusters of the Shapley core in section 4. In section 5,
the resulting two-dimensional distributions of X-ray surface
brightness, gas temperature, and metal abundance are discussed, and
finally, in section 6 a summary of our work is provided. Throughout
this paper we use the solar abundance table given by \citet{and89},
$H_0$ = 50 km s$^{-1}$ Mpc$^{-1}$, and $q_0$ = 0, and the quoted
errors of the spectral parameters are at the 90 \% confidence level.

\section{Observations}

Eight points in the core region of the Shapley Supercluster were
observed with $ASCA$. The longest observation among these pointings
was done for A3558, which is the brightest X-ray source among the five
targets in the core. In this paper we present the results of a
detailed study of this target. Data from the other four X-ray sources
are also analyzed with the same method. The observational log of these
data is shown in Table 1. \citet{hana99} and \citet{marke97} have
already reported the results of analyses for the core without the
additional observation of A3558 (observation 8 in Table 1).

We analyzed $ASCA$ data using LHEASOFT, version 5.0.1, supplied by
NASA Goddard Space Flight Center. After data screening with the
standard criterion, a rise-time discrimination technique was applied
for data of the Gas Imaging Spectrometer (GIS) to reject particle
events and a $sisclean$ for data of the Solid-State Imaging
Spectrometer (SIS) to exclude hot pixels. We estimated backgrouds from
point-sources-removed data of GIS blank-sky files\footnote{See
http://heasarc.gsfc.nasa.gov/docs/asca/mkgisbgd/mkgisbgd.html.} and
standard data (1994 November) of SIS blank-sky files.

\section{Results of \asca\ Data Analysis of A3558}

A3558, located at the center of the Shapley Supercluster core, has
been observed three times with $ASCA$ (observations 1, 4, and 8 in
Table 1). The total exposure time of these observations is 120 ksec
for each GIS detector, and the total number of photons within a radius
of 12$^\prime$ ($\sim$ 1 $h_{50}^{-1}$ Mpc) is 1.6 $\times$ 10$^5$
counts per detector.

Figure 2 shows an X-ray image obtained by these $ASCA$ GIS
observations. After the background subtraction, a correction of
surface brightness was done using the exposure map and GIS detector
efficiency map. The central X-ray peak in Figure 2 coincides with
A3558, and the east peak is a poor cluster, SC 1327$-$312. X-ray
energy spectra of GISs within a radius of 12$^\prime$ ($\sim$ 1
$h_{50}^{-1}$ Mpc) at the center of A3558 were evaluated with an
absorbed thermal (Raymond-Smith) model. The flux-weighted gas
temperature and abundance are estimated to be 5.59$\pm$0.06 keV and
0.34$\pm$0.02 solar, respectively. The 2--10 keV luminosity is
(6.8$\pm$0.5) $\times$ 10$^{44}$ $h_{50}^{-2}$ \lum. The redshift and
galactic hydrogen column density for absorption are fixed at 0.0482
\citep{abell89} and 3.89 $\times$ 10$^{20}$ cm$^{-2}$ \citep{dic90},
respectively. The spectral parameters obtained from $ASCA$ GIS and SIS
data are consistent within a 90 \% confidence level.

As seen in the $ROSAT$ image, $ASCA$ also shows that the X-ray surface
brightness distribution of A3558 is elongated in the
northwest--southeast direction, even in the higher energy band. This
structure is thought to be observational evidence that a cluster
merging has already happened in A3558 one or more times. In order to
confirm this scenario and investigate the physical processes in
detail, two-dimensional distributions of temperature and abundance
were made. These maps are powerful tools for understanding the
dynamical activity and evolution of the galaxy cluster.

\subsection{Assessment of the Merging Effect in Temperature and Abundance Maps}

The merging process distorts not only the distribution of the X-ray
surface brightness but also the distributions of temperature and
abundance. Some authors have already revealed azimuthally asymmetric
temperature maps, even for clusters with symmetric surface brightness
distribution, which had been thought to have been isolated for a long
time \citep{nabe99}. It takes typically several Gyrs to smooth the
X-ray surface brightness, although it depends on the mass ratio of the
clusters or the impact parameter of the merging. On the other hand,
the distributions of temperature and abundance can keep information
about the merging longer than that of surface brightness. Therefore,
temperature and abundance maps are very important for studying a
merger history. We derived the distributions of the temperature and
abundance of A3558 by using some hardness ratio maps. A detailed
description of the method is below.

A hardness ratio is an indicator of gas temperature. For an X-ray
cluster with a temperature of 5.59 keV, the hardness ratio of the
2.5--10 keV energy band (the hard band) to the 0.7--2.5 keV energy
band (the soft band) is $\sim$ 0.57 in the case of the $ASCA$ GIS. 
After background subtraction, a two-dimensional hardness ratio plot
was made by dividing the hard-band image by the soft-band image with
an angular resolution of 2$^\prime$ $\times$ 2$^\prime$. This map
shows that the gradient of the hardness ratio coincides with the
direction of the major axis of the X-ray surface brightness of A3558. 
The hardness ratio map is peaked at $\sim$ 4$^\prime$ northwest from
the central region of A3558 and decreases in the northwest and
southeast directions.

A two-dimensional temperature map is obtained from the hardness ratio
map in the following way: Assuming isothermal spherical symmetry of
the surface brightness, which is fitted by a single $\beta$-model with
parameters of a core radius of 4$^\prime$\hspace{-2pt}.\hspace{2pt}2
and $\beta$ of 0.611 from \citet{bard96}, hardness ratio maps are
simulated using an $ASCA$ full simulation program at a given
temperature from 3.5 keV to 7.5 keV with a step of 0.1 keV. From these
maps, the relation between the simulated hardness ratio and the
assumed temperature at each position on the detectors is derived. The
hardness ratio map is converted according to this relation into a
temperature map. An estimated temperature map within the radius of
12$^\prime$ at the center of A3558 is shown in Figure 3. The
superposed contour is an X-ray image in the 0.7--10 keV band,
corrected by the exposure map and the $ASCA$ GIS detector efficiency
map as shown in Figure 2. The error shown in this figure is the
largest error of the temperature at 90 \% confidence within the radius
of 12$^\prime$. This temperature map also shows the gradient along the
major axis of A3558 as seen in the hardness ratio map. The peak value
of the temperature map is 6.5 keV at the position of $\sim$ 4$^\prime$
northwest of the center of A3558, and the temperature decreases by
$\sim$ 1 keV farther northwest. This map also shows a possible
temperature drop in the southeast direction from the cluster center. 
Although these characteristics are consistent with the result shown by
\citet{marke97}, our derived temperature map enables us to investigate
spatial structure in detail. The derived temperature is ranging mainly
over 5.0--6.5 keV. No remarkable high-temperature regions are seen
within the radius of 12$^\prime$, such as the hot regions with
temperatures of over 10 keV seen in typical merging clusters, e.g.,
the 0.8 $h_{50}^{-1}$ Mpc off-center region of the Coma Cluster
\citep{nabe99} and the 1.0 $h_{50}^{-1}$ Mpc off-center region of the
Ophiuchus cluster \citep{nabe01}. Moreover, there are also no clear,
striking low-temperature regions, such as a cool region (cold front) 2
keV lower than the other regions in the merging cluster, e.g., A3667
\citep{vik01} and A2142 \citep{marke00}, which were revealed by
$Chandra$ observations.

Because gravitational settling of heavy elements takes much longer
than the age of the Universe \citep{sara88}, the observed abundance
map can constrain the amount of mixing due to a small-scale merging or
turbulence that can take place in a galaxy cluster. Moreover, even if
the large-scale cluster-cluster merging has occurred, their abundance
distributions have disturbed by the merging process less effectively
than their temperature maps. That is, not only the chemical evolution
but also the dynamical evolution can be revealed from an abundance
map. To get a two-dimensional distribution of metal abundance, an
intensity ratio of the iron K line and the neighboring continuum
emission is estimated. First, images in two energy bands, 5.0--6.0 keV
and 6.0--7.0 keV, are made from $ASCA$ data at the spatial resolution
of 4$^\prime$ $\times$ 4$^\prime$. The image of the 5.0--6.0 keV
energy band can be safely expected to include only continuum emission.
On the other hand, the 6.0--7.0 keV image includes iron lines in
addition to the continuum emission. Next, the two images of the
6.0--7.0 keV continuum emission and the 5.0--6.0 keV continuum
emission are simulated using the $ASCA$ full simulation program. In
this simulation, the temperature is assumed to be 6.75 keV, which is
obtained from $ASCA$ GIS and SIS 4--10 keV spectra within the radius
of 4$^\prime$. The assumed parameters of the $\beta$-model are the
same as the ones used for estimating temperature map \citep{bard96}. 
From these simulated images, the two-dimensional continuum intensity
ratio map, $f$ = $I_{6-7keV}$/$I_{5-6keV}$, is obtained. By
multiplying the $ASCA$ observed 5.0--6.0 keV image by this ratio map
$f$, we can guess the 6.0--7.0 keV continuum image. An iron-line
intensity map is derived by subtracting this estimated continuum image
from the $ASCA$ observed 6.0--7.0 keV image. Dividing this iron-line
map by the simulated 6.0--7.0 keV continuum image gives a ratio map of
iron-line intensity to continuum, corresponding to an equivalent-width
map. Finally, an abundance map is obtained by normalizing the value
averaged within a circle of radius 4$^\prime$. The obtained abundance
map is shown in Figure 4. Although the above described simulation was
done on the assumption of isothermal distribution, the
non-isothermality mentioned in this section causes an uncertainty of
the abundance of $\sim$ 10 \%.

In Figure 4, 4 pixels within the southeast region 12$^\prime$ away
from the center show high abundance, and this is also implied in the
south region at the same distance form the center, while other regions
are almost the same within the error.

\subsection{Detailed Structure of X-ray Surface Brightness Distribution}

To know the relation of the X-ray surface brightness distribution to
some interesting features seen in the temperature and abundance maps,
and moreover to get higher order structures of the surface brightness
distribution, we investigated the deviation of the X-ray surface
brightness from the smooth function described with an elliptical
$\beta$-model. The 0.7--10 keV $ASCA$ image was simulated using the
$ASCA$ full simulation program, assuming a core radius of
4$^\prime$\hspace{-2pt}.\hspace{2pt}9 and
3$^\prime$\hspace{-2pt}.\hspace{2pt}6, a $\beta$ of 0.61
\citep{bard96}, a temperature of 5.8 keV and an abundance of 0.32
solar. The temperature and abundance were obtained from only the
$ASCA$ GIS spectrum within a radius of 4$^\prime$.

By subtracting the simulated image from the observed 0.7--10 keV
image, we got substructures as shown in Figure 5. This residual image
is shown in units of the brightness fluctuation of the observed
0.7--10 keV image, that is, $\sigma$. Excess emission over the 3
$\sigma$ level is seen in the central region and expands to the
southeast. On the other hand, there are two drops of over the 3
$\sigma$ level in regions $\sim$ 5$^\prime$ northwest and $\sim$
10$^\prime$ southeast along the spheroidal major axis. These
structures resemble the sharp edge shown in many merging clusters.
The central excess flux within a radius of
2$^\prime$\hspace{-2pt}.\hspace{2pt}5 is estimated to be $\sim$ 9 \%
of the whole cluster. Since there is a dominant galaxy at the center
of A3558, corresponding to the central excess region, it can be
thought that the excess flux is associated with this galaxy. If it is
so, the corresponding luminosity of this excess is estimated to be 1.4
$\times$ 10$^{43}$ $h_{50}^{-2}$ \lum. The spectrum of this region is
well fitted with a single-temperature thermal (Raymond-Smith) model
with a temperature of 5.8$\pm$0.1 keV and an abundance of
0.35$\pm$0.04 solar at the reduced $\chi^2$ of 1.17. Two temperature
model fitting has been done, and the obtained parameters are
temperatures of 6.4$_{-0.3}^{+0.2}$ keV and 2.0$_{-0.6}^{+1.2}$ keV,
and an abundance of 0.35$\pm$0.04 solar, although the reduced $\chi^2$
value is not improved significantly.

According to a comparison of this residual image with the temperature
map and the abundance map, no clear excess has appeared at any regions
showing peculiarities in the temperature and abundance maps. On the
contrary, in the hot region $\sim$ 4$^\prime$ northwest from the
cluster center, the surface brightness decreases. Distributions of
surface brightness and temperature in this region resemble those of
the cold front in A2142 and A3667 revealed by $Chandra$ observation
\citep{marke00, vik01}, which is thought to be a boundary of the
remaining core of the merger component, moving through the hotter
ambient intracluster medium. Their hot faint regions have cold fronts
that lie close to the center of these clusters, because the core has
already passed through the central region of another merged cluster.

\subsection{Peculiar Regions within A3558}

As described in sections 3.1 and 3.2, peculiar regions in the spatial
distributions of temperature, abundance and residual X-ray surface
brightness are detected. To study these regions in detail by spectra,
we define three circular regions with a radius of 4$^\prime$
(corresponding to 320 $h_{50}^{-1}$ kpc ) named region 1, region 2,
and region 3. These regions are shown by blue circles in Figures 3, 4,
and 5. Their characteristics are a hot faint region, a bright region
with a southeast tail of central surface brightness excess, and a
high-abundance region, respectively. The central positions of these
peculiar regions are shown in Table 2. Each spectrum integrated in
these regions was fitted with an absorbed thermal (Raymond-Smith)
model, in order to confirm their characteristics of temperature and
abundance. Results are also shown in Table 2. The temperature of
region 1 and abundance of region 3 are 6.12$\pm$0.18 keV and
0.58$\pm$0.11 solar, respectively. Thus, region 1 indicates a
significantly high temperature and region 3 has a significantly high
abundance compared to the spectral parameters of the whole region.
According to the spectral fitting, it is noted that another
high-abundance region in the south $\sim$ 12$^\prime$ from the center
does not show a significantly high abundance. Therefore, we do not
focus on this region.

On region 3 we performed further analysis to investigate the origin of
the abundance excess. The redshift estimated from the center energy of
the iron K line is 0.047$_{-0.004}^{+0.010}$, which coincides with the
average value obtained for the whole region of A3558,
0.051$_{-0.002}^{+0.005}$, and the average redshift for member
galaxies, 0.0482 \citep{abell89}, within the error. Therefore, the
newly detected high-abundance region is likely to be within the same
gravitationally bounded system as A3558. In order to know the extent
of the metal-rich region, the abundances within the various integrated
radii were estimated. Figure 6 shows the dependence of the abundance
on the integrated radius. The region within a radius of 5$^\prime$ has
a significantly high abundance at a 90 \% confidence level. The
significantly metal-rich region is thought to be within a radius of
4$^\prime$, that is, 320 $h_{50}^{-1}$ kpc, considering the
point-spread function of the $ASCA$ X-ray Telescope. The spectrum of
region 3 is shown in the right panel of Figure 7. Because the
abundance estimation is influenced by temperature, doubts are casted
on the reliability of the high abundance. However, the temperature in
this region, 5.17$^{+0.27}_{-0.26}$ keV, is slightly lower than the
most effective iron-emitting temperature, 6.6 keV. Therefore, even if
the temperature is 6.6 keV, the abundance decreases to only 0.55
solar. Thus, the high abundance in this region is confirmed. It should
be noted that both spectra in Figure 7 show high-energy tails. 
Although these structures are thought to be due to additional
components, the spectral parameters do not suffer much from them.

It is thought to be clear that there is an additional metal-rich
component at the same redshift as A3558. The excess iron abundance
within a radius of 4$^\prime$ is 0.3$\pm$0.1 solar. However, because
excess brightness in the whole energy band of 0.7--10 keV is not seen
in Figure 5, it is thought that there is no significant additional gas
cloud. Then, the mean gas density of this region is estimated to be
$\sim$ 0.3 $\times$ 10$^{-3}$ $h_{50}^{0.5}$ cm$^{-3}$ on the
assumption of a $\beta$-model gas distribution for the main cluster
(core radius and $\beta$ referred from \citet{bard96}) with a central
electron density of 2.3 $\times$ 10$^{-3}$ $h_{50}^{0.5}$ cm$^{-3}$. 
The gas mass and the excess iron mass of the sphere within a radius of
320 $h_{50}^{-1}$ kpc are estimated to be 8 $\times$ 10$^{11}$
$h_{50}^{-2.5}$ M$_{\odot}$ and 5 $\times$ 10$^{9}$ $h_{50}^{-2.5}$
M$_{\odot}$, respectively.

\section{Five Galaxy Clusters in the Core Region of the Shapley Supercluster}

All $ASCA$ data of A3562, SC 1329$-$313, SC 1327$-$312, and A3556 were
analyzed with the same method as A3558. The whole obtained image of
the core region of the supercluster is shown in Figure 1. Not all
positions of the X-ray peaks of these five clusters are coincident
with Figure 1 of \citet{hana99} (the discrepancy of the peak position
for SC 1327$-$312 is over 4$^\prime$) but are consistent with the
positions obtained from galaxy distribution (from the NASA/IPAC
Extragalactic Database [NED]) within 1$^\prime$. The temperature maps
were not investigated by \citet{hana99}, except for SC 1329$-$313.
Therefore, we estimated the maps for all member clusters with
4$^\prime$ $\times$ 4$^\prime$ pixels. Three Abell clusters in the
core region have asymmetric spatial distributions of the gas
temperature, as shown in the hardness ratio map of Figure 8. For A3556
and A3562, the hardness ratio at the side toward A3558 seems to be
higher than that toward the opposite side. Practically, spectra of the
east and west sides of A3556 within a
12$^\prime$\hspace{-2pt}.\hspace{2pt}5 radius show that the
temperature of the east side, 3.8$^{+1.1}_{-0.7}$ keV, is higher than
that of the west side, 2.6$^{+0.5}_{-0.4}$ keV. The spectrum of the
southwest region of A3562 within a radius of
12$^\prime$\hspace{-2pt}.\hspace{2pt}9 shows a temperature of
5.9$_{-0.7}^{+1.1}$ keV, which tends to be higher than that of the
northeast region, 4.6$^{+0.8}_{-0.6}$ keV, although it is not
significant. It is thought to be due to a cluster-cluster merging or a
group infalling. On the other hand, the two poor clusters show
relatively isotropic temperature distributions on the scale of $\sim$
0.3 $h_{50}^{-1}$ Mpc.

Gas temperature, abundance, and luminosity within a radius of
12$^\prime$ for each cluster in the Shapley core are shown in Table 3. 
The total gravitational and gas mass of this core region can be
estimated from these results and $\beta$-model parameters from those
published \citep{bard96, ettori97, ettori00}. The total gas mass is
obtained to be $\sim$ 2 $\times$ 10$^{14}$ $h_{50}^{-2.5}$ M$_{\odot}$
and the total gravitational mass is $\sim$ 2 $\times$ 10$^{15}$
$h_{50}^{-1}$ M$_{\odot}$. Therefore, the mean gravitational density
within a radius of 6.5 $h_{50}^{-1}$ Mpc is 25 times larger than the
critical density $\rho_c$.

Both the luminosity and temperature relation and the gravitational
mass and temperature relation for these five clusters are consistent
with those of the nearby clusters with a redshift of less than 0.1
investigated by \citet{aki01} and \citet{matsu00} within the
deviations of their data.  This result suggests that these five
clusters are not peculiar clusters of galaxies.

\section{Discussion}

The two-dimensional distributions of surface brightness, temperature
and abundance of A3558 indicate this cluster has a high dynamical
activity. This result is expected from the situation that this cluster
resides at the center of a supercluster core with a high number
density of member clusters. According to the distributions we
investigated, this cluster is most probably suggested to at least have
experienced mergings twice in the past and possibly to have had a
recent group infalling.

The X-ray surface brightness distribution of A3558, showing elongation
in the northwest and southeast direction, suggests that a merging of
two large galaxy clusters with equal mass (major merger) has taken
place along the major axis in the past, according to the results of an
N-body simulation of head-on mergers \citep{roet97}. Although the
spatial distribution of the temperature of A3558 is asymmetric, a hot
region 3$^\prime$ away from the cluster center northeastward does not
spread out much, and the difference of temperature at each position
within A3558 is small. Therefore, we think that it has been $\sim$ 5
Gyr since this major merging \citep{roet97}. In the case of the Coma
Cluster, there is a large temperature fluctuation, even though it
shows a smooth surface brightness \citep{nabe99}. This focusing
cluster is a contrary case to the Coma Cluster. It is not clear
whether this merger had a small impact parameter or a large one. An
N-body simulation of an off-set merger for A3266 \citep{roet00} also
gives a distribution of surface brightness that resembles A3558. 
Therefore, a scenario that a merging with a large impact parameter has
taken place along the major axis in the past is not able to be
rejected.

The characteristics of temperature and surface brightness in region 1
are discussed. They look like a cold front due to cluster merging
reported by some authors based on $Chandra$ observations
(\citet{marke00},\citet{vik01} and so on). Considering the passage of
time from the merging, it is difficult to think that these small-scale
structures could have been formed by the above-described large-scale
merger and still survive. It is plausible that after the large-scale
merger occurred, the main cluster and a subcluster infalling from the
northwest merged with a small impact parameter, and an area similar to
a cold front was formed; however no striking high- or low-temperature
regions were made. According to the simulation of offset mergings with
a mass ratio of 1 : 4 by \citet{taki00}, a bow shock and a merged cold
core are seen for 0.25 Gyr after the closest contraction epoch of two
merging clusters. At the end of the simulated time span (0.7 Gyr),
although the distribution of surface brightness is still asymmetric,
the fluctuation of temperature disappears. Therefore, it is thought
that only less than 0.7 Gyr has passed after this small-scale merger. 
It is noted that the large elongated gas profile of A3558 would not
have formed within 0.7 Gyr after the merging of one small-scale merger
such as this, according to comparison in the case of various mass
ratios presented by \citet{taki00} and \citet{roet97}. The large
ellipticity is likely to be due to the large-scale merging (major
merger) mentioned above. Thus, the result that two cluster mergings
happened is beyond a doubt.

On the origin of the high-abundance region (region 3), the most
probable one of some possibilities is that it is a remnant of a merged
subcluster core, which often shows abundance enhancement at an evolved
stage. The diffusion time of iron is more than 15 Gyr. Therefore, the
iron can persist for a long period in the remnant core after any
dynamical event. Gas density fluctuations, on the other hand, fade
away on the sound-crossing timescale, and any excess gas density is
smeared out. In this way, regions with a typical size of a few hundred
kiloparsecs with a significant iron excess but without a noticable
density excess can form.

In this cluster, the second small-scale merging with the passage of
time of less than 0.7 Gyr cannot make the remnant of a merged core
because this region is far from the cluster center. On the other hand,
it is thought that the first major merging can form the iron remnant. 
After the first major merging, only the iron core can survive, since
the diffusion time of iron is longer than $\sim$ 5 Gyr, that is, the
passage of time after the large-scale merging.

To examine our senario of a remnant of a merged cluster core with high
metalicity, we constructed a sinple toy model assuming an initial size
of the merged cluster of 300 $h_{50}^{-1}$ kpc, an abundance of 0.58
solar, and a central density of several 10$^{-3}$ cm$^{-3}$, which is
10 times larger than the density of the ambient intracluster medium. 
After 0.3 Gyr, the radius of the core spreads at the sound speed of
1150 km s$^{-1}$ and reaches 650 $h_{50}^{-1}$ kpc. The gas density of
the core comes to be the same value as that of the ambient
intracluster medium, and then the brightness excess is smeared. The
core size of iron ions, however, can increase to only 20 $h_{50}^{-1}$
kpc within 0.3 Gyr, since iron ions expand through random walk with a
thermal velocity of 130 km s$^{-1}$. The size of the spread can be
described by the product of the mean free path and square root of the
number of collisions. The averaged number of collisions is 4.4. Thus,
a region with no electron density excess, that is, no surface
brightness excess, but with a high metal abundance will be able to be
formed. After the formation, large-scale merging and mixing may have
not been induced. Nonetheless, thermal conduction, dynamic viscosity,
ram-pressure stripping, and Rayleigh-Taylor and Kelvin-Helmholtz
instabilities make the time for an iron core to survive shorter,
unless these processes are suppressed by magnetic fields.

Another suggestion is the existence of an active galactic nucleus
(AGN) such as a type II Seyfert galaxy. The spatial distribution
cannot determine whether the residual structure is caused by a point
source or not, but the line center energy of iron K$\alpha$, that is,
the redshift is coincident within the error with the mean value of the
overall region of the cluster within a radius of 12$^\prime$, and the
time variability of this region is not detected significantly. No
point source has been found in images of $ROSAT$, and no nonthermal
component is needed to describe the $ASCA$ spectra. These facts do not
support the idea of an AGN.

The other suggestion is the existence of a metal-rich group of
galaxies. If there is a galaxy group that is binding the gas within
320 $h_{50}^{-1}$ kpc, it means that this region does not show strong
interactions with A3558, and this idea is able to explain that the
size of 320 $h_{50}^{-1}$ kpc is the typical size of groups of
galaxies. If there is a group of galaxies, it has a gas mass of 8
$\times$ 10$^{11}$ $h_{50}^{-2.5}$ M$_{\odot}$ and an excess iron mass
of 5 $\times$ 10$^9$ $h_{50}^{-2.5}$ M$_{\odot}$ within a sphere of
this radius. The gas mass was derived assuming that the sphere has a
uniform density. Bcecause there is no clear electron density excess,
the electron number density was set to be 0.0003 cm$^{-3}$ at the
center of this region. Thus, the gas mass was obtained. On the other
hand, the gas mass within a column with a projected radius of
4$^\prime$ is 7 $\times$ 10$^{12}$ $h_{50}^{-2.5}$ M$_{\odot}$. 
Therefore, the iron mass is 1 $\times$ 10$^{10}$ $h_{50}^{-2.5}$
M$_{\odot}$ for an abundance of 0.58. The excess iron mass is 41 \% of
the iron mass, that is, 5 $\times$ 10$^9$ $h_{50}^{-2.5}$ M$_{\odot}$.
If the abundance for other regions of this sphere is 0.34, the sphere
is then though to have excess iron mass. These values are reasonable
for a group of galaxies. For example, the radii of the X-ray extent of
78 groups and the intragroup medium masses of 24 groups are shown in
\citet{mul96}. The iron abundance of galaxy groups varies
significantly from group to group: 0 -- 1 solar metallicity
\citep{ren97}; therefore, the iron mass also varies between $\sim$
10$^8$ and $\sim$ 10$^9$ $h_{50}^{-2.5}$ M$_{\odot}$.

In this paper, the iron K line is used, which is advantageous. There
are 38 galaxies within the projected radius of 4$^\prime$ (320
$h_{50}^{-1}$ kpc) according to the NED. Since not all of their
redshifts are known, it is assumed that all are within this region. 
Because their total $B$-band luminosity is 2.3 $\times$ 10$^{11}$
L$_{\rm B}$$_\odot$, the luminous mass is 2 $\times$ 10$^{12}$
M$_{\odot}$ on the assumption that the mass-to-luminosity ratio is 8
for an elliptical galaxy. The ratio of the iron mass to the total
optical luminosity of the galaxies is obtained to be 0.02 M$_{\odot}$
L$_{\rm B}$$_\odot$$^{-1}$, which is also acceptable for galaxy
groups. The ratio of galaxy groups varies 0.0001--0.02 M$_{\odot}$
L$_{\rm B}$$_\odot$$^{-1}$ \citep{ren97}. Because not all of the 38
galaxies are members of the group, the total $B$-band luminosity and
luminous mass are upper limits, and the ratio is a lower limit. 
Therefore, several other possible interpretations cannot be ruled out,
such as bursts of star formation occurring because of group infalling,
stars ejecting a large amount of iron into the intragroup medium and
then disappearing, or dust grains with condensed iron being
evaporated.

Finally, the mean density of the core region in the Shapley
Supercluster is 25 times larger than the critical density. Although it
is much smaller than needed to virialize, that is, 178 times critical
density, it is larger than that needed to decouple from the cosmic
expansion and contract, that is, 5.6 times critical density
\citep{peebles80}. Therefore, the core region of the Shapley
Supercluster is on the way to contraction and is collapsing to form a
virialized halo. However, because the characteristics of each member
cluster in the core region are consistent with other typical nearby
galaxy clusters, the core region of the Shapley Supercluster is
probably in the early phase of contraction. This scenario is supported
by the fact that the core region is plotted at the right edge (the
largest scale radius $r_s$ and the lowest characteristic density
$\delta_{c}$) in Figure 4 of \citet{sato99}.

\section{Conclusions}

The results of the $ASCA$ observations of five member clusters in the
core region of the Shapley Supercluster, mainly about A3558, are
shown. The X-ray surface brightness distribution of A3558 is
elongated in the northwest and southeast directions and is furthermore
asymmetric to the axis. The temperature map shows that the central
region within a radius of 1 $h_{50}^{-1}$ Mpc has a temperature
gradient along the major axis. The range of temperature is 5.0--6.5
keV, and there are not extremely hot or cool regions. The obtained
abundance map shows that the range of abundance is 0.3--0.5 solar and
nearly isotropic. Spatial distributions of X-ray surface brightness,
temperature, and abundance indicate a hot depression region $\sim$
4$^\prime$ northwest (region 1), X-ray excess $\sim$ 4$^\prime$
southeast (region 2), and a metal-rich region within $\sim$ 320
$h_{50}^{-1}$ kpc, 12$^\prime$ southeast from the cluster center of
A3558 (region 3).

The overall temperature and elongated distributions of surface
brightness are thought to have arisen from a large scale merging of
two large galaxy clusters that occurred $\sim$ 5 Gyr ago. The hot
depression region (region 1) implies that a small-scale merging with
the main cluster and a subcluster infalling from the northwest
occurred less than 0.7 Gyr ago and after a large-scale merging. The
anisotropic distributions of surface brightness and temperature
indicate that mass infall along the major axis is dominant in the core
region of the Shapley Supercluster. As concerns the metal-rich region
(region 3), it is probably the remnant of a core region in a galaxy
cluster that merged $\sim$ 5 Gyr ago. If it is not a remnant, it is
possible that an infalling group of galaxies binding X-ray-emitting
gas within 320 $h_{50}^{-1}$ kpc exists in this region. Thus, two
mergers and a possible infalling of a galaxy group are suggested. It
is thought that this cluster has a high dynamical activity due to the
particular environment in the supercluster core.

Moreover, the gas temperature, abundance, luminosity, gas mass, and
gravitational mass of the other four member clusters belonging in core
region of the Shapley Supercluster have been estimated. The values of
these five galaxy clusters show the same correlations of $L_{\rm
X}-kT$ and $M_{\rm gas}-kT$ relations as nearby galaxy clusters with a
redshift of less than 0.1. Even though these five clusters of galaxies
are in the highest density circumstance and some clusters show an
anisotropic temperature distribution, they have no differences with
field clusters of galaxies; that is, the contribution of each
interaction of the member clusters to overall spectral parameters is
thought to be weak. The sum of the gravitational mass of the five
galaxy clusters is $\sim$ 2 $\times$ 10$^{15}$ M$_{\odot}$. There are
no more significant X-ray sources bounding dark matter within the core
region. Therefore, the mean density of the core region of the Shapley
Supercluster is about 25 times the critical density, which is smaller
than 178 to start virializing but is larger than 5.5 to start
contracting. The core region is just on the way to the contraction.

In order to make our discussion clear, observations with $Chandra$,
$XMM$, and $ASTRO-E$ II, which will be launched in 2005, are needed.

\acknowledgments

We would like to thank all the members of the \asca\ team for their
operation of the satellite.  We are grateful to the referee for useful
suggestions and valuable comments.\\

\clearpage

\figcaption[f1.ps]
{
$ASCA$ image of the core region in the Shapley Supercluster. The
energy band is 0.7--10 keV. The circles show $ASCA$ fields of view. 
The adopted equinox year is 2000.
}
\figcaption[f2.ps]
{
$ASCA$ GIS image for the 0.7--10 keV energy band from observations 1,
4, and 8 in Table 1. The radius of the circle is 1 $h_{50}^{-1}$ Mpc
(see text for details).
}
\figcaption[f3.ps]
{
$ASCA$ temperature map of A3558. The unita are keV. The radius of the
red circle is 1 $h_{50}^{-1}$ Mpc ($=$ 12$^\prime$), and the dashed
line shows the major axis. The superposed contours show the X-ray
surface brightness of the 0.7--10 keV energy band. The pixel size is
2$^\prime$ $\times$ 2$^\prime$. The error shown by the arrow is the
maximum error of the temperature within a radius of 12$^\prime$. The
blue circle shows region 1.
}
\figcaption[f4.ps]
{
$ASCA$ abundance map of A3558. The units are in solar metallicity. The
red circle and superposed contour are the same as for Fig. 3. The
error shown by the arrow is the maximum error of the abundance within
a radius of 12$^\prime$. The pixel size is 4$^\prime$ $\times$
4$^\prime$. The blue circle shows region 3.
}
\figcaption[f5.ps]
{
$ASCA$ residual image of A3558. The units are in $\sigma$. The dashed
line, red circle, and superposed contour are the same as for Fig. 3. 
The blue circle shows region 2.
}
\figcaption[f6.ps]
{
Relation between the integrated radius and the abundance of region 3. 
The line shows the upper limit of the average abundance of A3558.
}
\figcaption[f7.ps]
{
$ASCA$ GIS spectra of the central region and high-abundance region
within a radius of 4$^\prime$. The temperature (keV) and abundance
(solar) obtained by fittings with a thermal (Raymond-Smith) model are
shown. $Right$: Spectrum of region 3. $Left$: Spectrum of the central
region shown for comparison.
}
\clearpage

\figcaption[f8.ps]
{
$ASCA$ hardness ratio map of the core region in the Shapley
Supercluster.  The hardness ratio is defined as the ratio of the
2.5--10 keV flux to the 0.7--2.5 keV flux. The superposed contours
show the X-ray surface brightness for 0.7--10 keV and are the same as
Fig. 1. The pixel size is 4$^\prime$ $\times$ 4$^\prime$.
}

\clearpage

\begin{table} 
\begin{center}
\caption{Log of $ASCA$ Observations}
\begin{tabular}{l l l c c c c} \tableline\tableline
No. & Target Name & Observed Date & \multicolumn{2}{c}{Exposure(ksec)} &
\multicolumn{2}{c}{Position Coordinate(J2000)} \\
 & & & GIS & SIS\tablenotemark{a} & $\alpha$ & $\delta$ \\ \tableline
1 & A3558 & 1994 07 14$\sim$15 & 20.8 & 20.4$^{\spadesuit}$ & 
$13^{\rm h}27^{\rm m}56^{\rm s}\hspace{-5pt}.\hspace{2pt}9$ & $-31^{\circ}29^{\prime}38^{\prime\prime}\hspace{-5pt}.\hspace{2pt}4$ \\
2 & A3556 & 1995 01 23$\sim$24 & 39.1 & 39.9$^{\spadesuit}$ & 
$13^{\rm h}24^{\rm m}06^{\rm s}\hspace{-5pt}.\hspace{2pt}0$ & $-31^{\circ}38^{\prime}60^{\prime\prime}\hspace{-5pt}.\hspace{2pt}0$ \\
3 & SC 1327$-$312 & 1995 01 24$\sim$25 & 30.4 & 29.4$^{\diamondsuit}$
& $13^{\rm h}29^{\rm m}45^{\rm s}\hspace{-5pt}.\hspace{2pt}4$ & $-31^{\circ}36^{\prime}11^{\prime\prime}\hspace{-5pt}.\hspace{2pt}9$ \\
4 & IC region\tablenotemark{b} & 
1995 07 23$\sim$24 & 31.9 & 30.8$^{\diamondsuit}$ & $13^{\rm h}29^{\rm m}00^{\rm s}\hspace{-5pt}.\hspace{2pt}0$ & 
$-31^{\circ}33^{\prime}50^{\prime\prime}\hspace{-5pt}.\hspace{2pt}0$ \\ 
5 & A3562 & 1996 07 27 & 19.8 & 20.2$^{\diamondsuit}$ & $13^{\rm h}33^{\rm m}31^{\rm s}\hspace{-5pt}.\hspace{2pt}8$ & 
$-31^{\circ}39^{\prime}36^{\prime\prime}\hspace{-5pt}.\hspace{2pt}0$\\
6 & SC region\tablenotemark{c}
& 1996 07 27$\sim$29 & 45.0 & 
44.9$^{\diamondsuit}$ & $13^{\rm h}31^{\rm m}52^{\rm s}\hspace{-5pt}.\hspace{2pt}8$ & $-31^{\circ}30^{\prime}25^{\prime\prime}\hspace{-5pt}.\hspace{2pt}2$ \\
7 & SC 1329$-$313 & 1996 07 29 & 27.7 & 28.5$^{\diamondsuit}$ & 
$13^{\rm h}31^{\rm m}35^{\rm s}\hspace{-5pt}.\hspace{2pt}9$ & $-31^{\circ}48^{\prime}44^{\prime\prime}\hspace{-5pt}.\hspace{2pt}6$ \\
8 & A3558 & 2000 01 21$\sim$24 & 71.7 & 66.5$^{\clubsuit}$ & 
$13^{\rm h}27^{\rm m}54^{\rm s}\hspace{-5pt}.\hspace{2pt}8$ & $-31^{\circ}29^{\prime}31^{\prime\prime}\hspace{-5pt}.\hspace{2pt}9$ \\ \tableline
\end{tabular}
\tablenotetext{}{
Notes. -- Units of right ascension are hours, minutes, and seconds,
and units of declination are degrees, arcminutes, and arcseconds. We
used three pointing data from observations 1, 4, and 8 for our
investigation of A3558.}
\tablenotetext{a}{Here spade ($\spadesuit$) symbols are for the 4CCD mode, diamonds ($\diamondsuit$) for the 2CCD mode, and clubs ($\clubsuit$) for the 1CCD mode.}
\tablenotetext{b}{The midpoint between A3558 and SC 1327$-$312}
\tablenotetext{c}{Off-center region near A3562, SC 1329$-$313, and
 SC 1327$-$312.} 
\end{center}
\end{table}

\begin{table}
\begin{center}
\caption{position and X-ray spectral parameters}
\begin{tabular}{l c c c c c} \tableline \tableline
Region No. & \multicolumn{2}{c}{position} & characteristic &  \multicolumn{2}{c}{spectral parameters} \\
 & $\alpha_{{\rm J}2000}$ & $\delta_{{\rm J}2000}$ &  & $kT$(keV) & Abundance(solar) \\ \tableline
1 & $13^{\rm h} 27^{\rm m} 46^{\rm s}\hspace{-5pt}.\hspace{2pt}4$ & $-31^{\circ} 26^{\prime} 03^{\prime \prime}$ & high temperature, depression & 6.12$\pm$0.18 & 0.32$\pm$0.04 \\
2 & $13^{\rm h} 28^{\rm m} 05^{\rm s}\hspace{-5pt}.\hspace{2pt}5$ & $-31^{\circ} 33^{\prime} 02^{\prime \prime}$ & excess emission & 5.51$^{+0.10}_{-0.09}$ & 0.33$\pm$0.03 \\
3 & $13^{\rm h} 28^{\rm m} 37^{\rm s}\hspace{-5pt}.\hspace{2pt}8$ & $-31^{\circ} 37^{\prime} 27^{\prime \prime}$ & metal rich & 5.17$^{+0.27}_{-0.26}$ & 0.58$\pm$0.11 \\
\tableline \tableline
\end{tabular}
\tablenotetext{}{
Notes. -- Units of right ascension are hours, minutes, and seconds,
and units of declination are degrees, arcminutes, and arcseconds.  The
radii of these regions are 4$^\prime$, and their redshift and
absorption are fixed at 0.0482 and 3.89$\times$10$^{20}$ cm$^{-2}$,
respectively.}
\end{center}
\end{table}

\begin{table}
\begin{center}
\caption{X-ray spectral parameters of five galaxy clusters}
\begin{tabular}{l c c c c c} \tableline \tableline
 & A3562 & SC 1329$-$313 & SC 1327$-$312 & A3558 & A3556 \\ \tableline
Redshift & 0.0490 & 0.0482 & 0.0495 & 0.0482 & 0.0479 \\
$kT$(keV) & 5.04$^{+0.23}_{-0.22}$ & 4.22$^{+0.25}_{-0.23}$ &
3.65$^{+0.14}_{-0.11}$ & 5.59$\pm$0.06 & 3.21$^{+0.25}_{-0.24}$ \\
Abundance(solar) & 0.40$\pm$0.08 & 0.24$^{+0.11}_{-0.10}$ & 
0.32$\pm$0.07 & 0.34$\pm$0.02 & 0.38$^{+0.21}_{-0.18}$ \\
$L_{\rm X}$($10^{44}$ergs s$^{-1}$) & 2.86$^{+0.86}_{-0.07}$ & 
0.92$^{+0.28}_{-0.03}$ & 1.66$^{+0.50}_{-0.04}$ & 6.83$^{+2.05}_{-0.04}$ & 
0.33$^{+0.10}_{-0.03}$ \\ \tableline \tableline
\end{tabular}
\tablenotetext{}{
Note. -- These are estimated from 0.7--10 keV spectra within a radius
of 12$^{\prime}$ ($\sim$ 1 $h_{50}^{-1}$ Mpc) of each target. The
$L_{\rm X}$ is the luminosity of the 2--10 keV band.
}
\end{center}
\end{table}

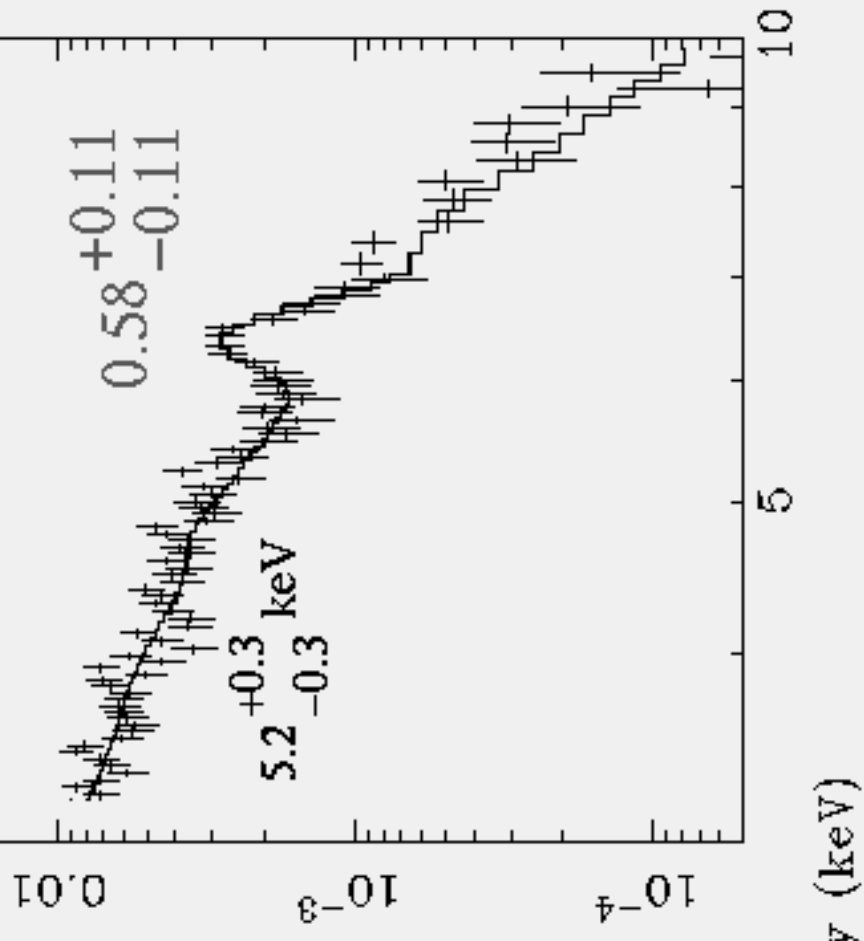
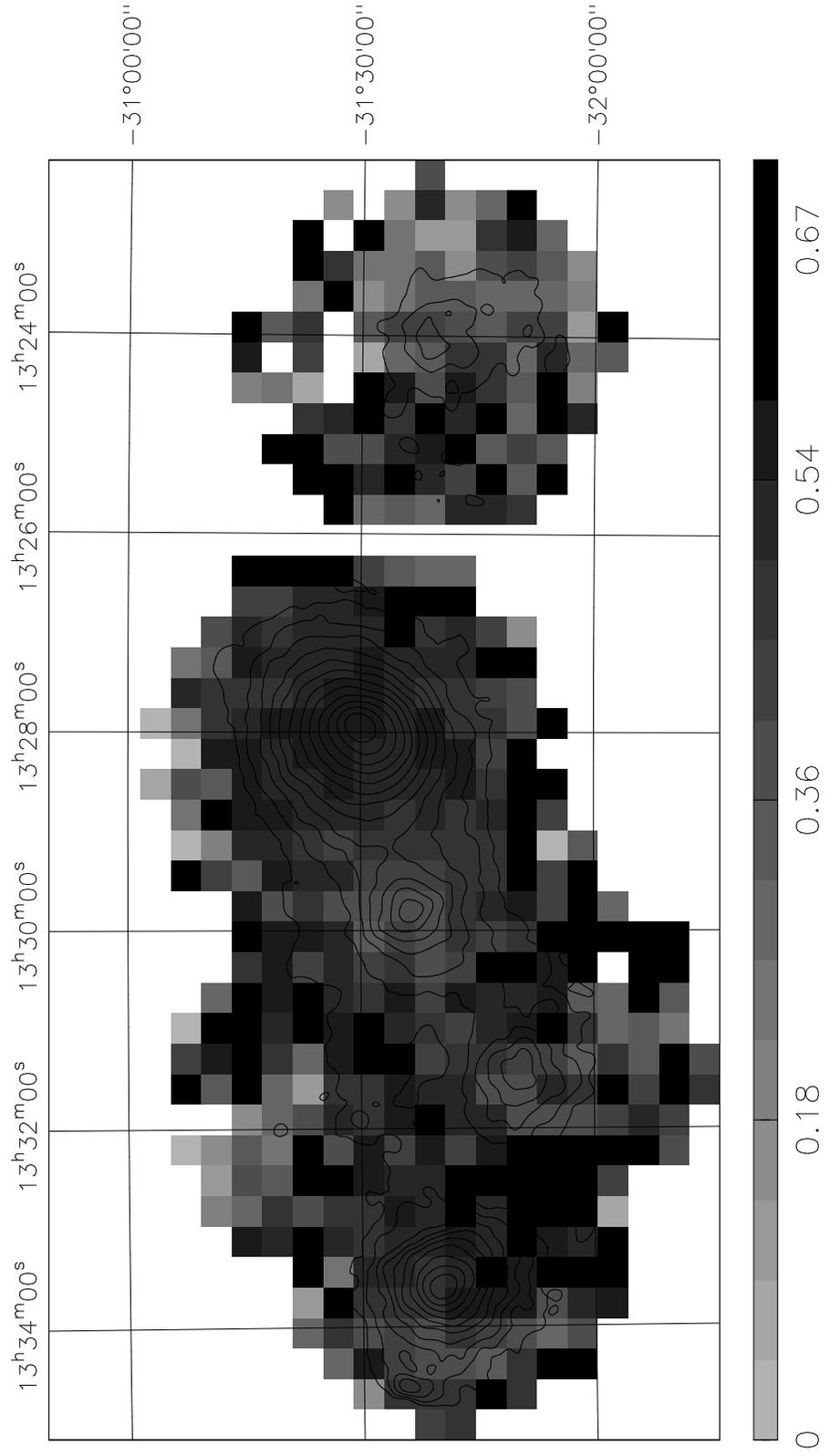
\end{document}